# Visible octave frequency combs in silicon nitride nanophotonic waveguides driven by Ti:sapphire lasers

**Abdullah Alabbadi**[1,2,3,*], **Ewan Allan**[4], **Hanna Ostapenko**[4], **Pablo Castro-Marín**[4], **Derryck T. Reid**[4], **And Pascal Del'Haye**[1,2,*]

[1]*Max Planck Institute for the Science of Light, 91058 Erlangen, Germany.*
[2]*Department of Physics, Friedrich-Alexander-Universität Erlangen-Nürnberg, 91058 Erlangen, Germany.*
[3]*Department of Physics, Faculty of Science, Alexandria University, Moharram Bek, 21511 Alexandria, Egypt*
[4]*Institute of Photonics and Quantum Sciences, School of Engineering and Physical Sciences, Heriot-Watt University, Edinburgh, EH14 4AS.*
*abdullah.alabbadi@mpl.mpg.de
*pascal.delhaye@mpl.mpg.de

**Abstract:**

Nonlinear nanophotonic waveguides have opened a route to compact frequency combs for precision metrology, spectroscopy and astronomy, yet broadband comb access to the visible remains challenging on CMOS-compatible platforms. Silicon nitride is widely accessible and low loss into the visible, but most demonstrations rely on telecom pumping and thick stress-managed films, where the large spectral gap to the visible dispersive wave raises the soliton order and power required for efficient conversion. Here we show that pumping closer to the visible provides a complementary route. Starting from crack-free 400 nm $Si_3N_4$ films, we implement dispersion-engineering with air-clad nanophotonic waveguides whose enhanced geometric dispersion opens an anomalous-dispersion window across the Ti:sapphire tuning range. Femtosecond Ti:sapphire pulses then drive octave-spanning combs from the visible to the near-infrared, with the visible edge and overall bandwidth lithographically tuned by the waveguide width at pulse energies of only tens of picojoules. The air-clad geometry also produces strong polarization-dependent dispersion, enabling switching between all-normal and soliton-dominated broadening in the same device, and support octave-spanning combs at 1 GHz repetition rates directly driven by a compact diode-pumped Ti:sapphire oscillator. These results position air-clad $Si_3N_4$ nanophotonic waveguides as an efficient interface between emerging short-wavelength integrated gain platforms and fully integrated visible frequency-comb engines.

Integrated nonlinear photonics has matured to the point where ultrafast lasers and nonlinear optics, once the domain of bulky optics and metre-scale fibers, can be sculpted lithographically onto a chip. Among CMOS-compatible platforms, silicon nitride ($Si_3N_4$) stands out for simultaneously offering large Kerr nonlinearity, low linear loss extending well into the visible, low two photon absorption, and fabrication routes compatible with industrial foundries[1,2]. High refractive-index contrast with silicon dioxide enables tight optical confinement in sub-micron cores, boosting the effective nonlinearity; on the other hand, precise lithographic control of dimensions allows for dispersion and modal confinement to be engineered with nanometer-scale accuracy[3,4]. That combination of high nonlinearity and lithographically controlled dispersion makes $Si_3N_4$ an excellent candidate for chip-scale frequency combs and supercontinuum generation[5].

Broadband visible supercontinua and frequency combs have a wide range of scientific and industrial applications. They underpin sensing of atmospheric pollutants[6], integrated visible-light biosensing and neurophotonics[7], quantum and nonlinear spectroscopy[8], and chip-based optical coherence tomography[9,10]. In astrophotonics, visible combs with wide and / or flexible GHz-scale mode spacings are central to extreme-precision radial-velocity measurements and calibration of high-resolution spectrographs in the search for exoplanets[11–14]. For many of these applications, especially optical frequency metrology and astrophotonics, octave-spanning bandwidth is particularly valuable as it enables self-referencing using f-2f schemes, simplifying carrier-envelope offset stabilization and directly linking widely separated optical transitions in the visible and near-infrared[15,16]. Delivering octave coverage in the visible-to-near-infrared on a CMOS-compatible chip is therefore technologically and scientifically very attractive.

The dynamics of supercontinuum generation in nonlinear waveguides are governed by a delicate interplay between the Kerr nonlinearity and the dispersion landscape[17]. In the time domain, high-peak-power pulses accumulate a nonlinear phase of $\phi_{NL(t)} = \gamma P(t) L_{eff}$, leading to self-phase modulation (SPM) that generates new frequencies and imparts an up-chirp on the pulse. In an all-normal-dispersion landscape ($\beta_2 > 0$), both SPM and group-velocity dispersion (GVD) contribute the same sign of chirp, so the pulse stretches temporally and the peak power decreases as the spectrum broadens. This regime can yield flat, highly coherent spectra that are compressible to few-cycle pulses, but typically requires relatively high pulse energies and modest normal dispersion to sustain large broadening[18–20]. In contrast, when the pump lies in the anomalous-dispersion regime ($\beta_2 < 0$), dispersion induces a down-chirp while SPM still produces an up-chirp. The competition between these contributions leads to soliton dynamics: higher-order solitons self-compress to extreme peak powers[21–23] and fission into fundamental solitons that, in the presence of higher-order dispersion, phase-match radiation into dispersive waves (DWs) in the normal-dispersion bands. These DWs inherit the coherence and comb structure of the pump, and their wavelengths determine the short- and long-wavelength edges of the continuum[24]. Efficient, coherent octave generation therefore requires both sufficiently strong nonlinearity and a dispersion landscape that supports bright solitons at the pump and phase-matched DWs spanning the target band.

Achieving such soliton-driven dynamics and dispersive wave generation at short pump wavelengths on $Si_3N_4$ is fundamentally challenging. The material dispersion of both $Si_3N_4$ and $SiO_2$ is strongly normal below ∼1.3 µm, so buried $Si_3N_4$ waveguides clad with $SiO_2$ above and below typically exhibit all-normal GVD throughout the visible and near-infrared. This regime is well suited for coherent and flat broadening[20] but is inefficient for reaching very broad bandwidth at low pulse energies. Consequently, integrated $Si_3N_4$ supercontinuum sources have predominantly been driven at telecom wavelengths[10,25], where the material dispersion is weaker and anomalous GVD can be obtained using thicker $Si_3N_4$ layers above the crack-free limit of

~400 nm[26]. This in turn requires sophisticated strategies to mitigate the high tensile stress and cracking in thick films, including stress-relief trenches, Damascene-style trench filling, and multilayer deposition,[26–30] which limit accessibility and complicate large-scale integration. Telecom pumping has been highly successful for octave spanning and mid-infrared extension in silicon nitride[24,31], but it is not an efficient starting point for broadband visible generation. Dispersive-wave emission into the visible must traverse a wide normal-dispersion interval and a large frequency separation from the soliton, which increases the soliton order and peak power required for strong spectral overlap and efficient visible transfer. As a result, telecom-driven spectra that reach the visible often do so with comparatively reduced visible spectral density, even when total bandwidth is large.

Pumping closer to the visible offers a fundamentally different route. At shorter pump wavelengths $\lambda$, the spectral separation and group-velocity mismatch between the pump and a target visible dispersive wave are reduced, lowering the phase-mismatch barrier and relaxing the required higher-order dispersion engineering[32,33]. In addition, the nonlinear parameter $\gamma$ of an integrated waveguide scales with wavelength as $\gamma(\lambda) = \frac{2\pi}{\lambda} \frac{n_2}{A_{eff}(\lambda)}$, where $n_2$ is the Kerr coefficient and $A_{eff}$ the effective mode area. For high-index-contrast strip waveguides, full-vector simulations show that the effective mode area optimum for high confinement decreases quadratically with wavelength[34], which results in an optimum nonlinear parameter approximated as:

$$\gamma(\lambda)_{opt} \propto \frac{n_2}{\lambda^3} (n_{core} + n_{clad})^{1.2}(n_{core} - n_{clad})^{0.8},$$

where $n_{core}$ and $n_{clad}$ are refractive index of waveguide core and cladding, respectively. Thus, shorter wavelengths and higher index contrast both enable tighter confinement in smaller cores, substantially boosting $\gamma$. Si$_3$N$_4$ has a Kerr coefficient of $n_2 \approx 2.45 \times 10^{-19}$ m$^2$/W, roughly an order of magnitude higher than silica,[35] and when combined with the tight mode area in a sub-µm core at short wavelength, the nonlinear parameter $\gamma$ can reach ~10 W/m at 800 nm wavelength, which is three order of magnitude larger than the nonlinear parameter in highly nonlinear fibers[36]. This makes nonlinear processes such as supercontinuum generation intrinsically more efficient when driven at 800 nm rather than 1.55 µm, provided anomalous GVD can be realized.

Ti:sapphire lasers are uniquely suited to exploit this regime. Their exceptionally broad gain bandwidth, spanning roughly from 650 nm to 1050 nm, supports sub-10-fs pulses and wide tunability. As a result, they have been the workhorse for ultrafast science and visible-near-infrared frequency combs for decades[16]. Recent breakthroughs in Ti:sapphire photonic integration have brought this performance toward the chip scale[37–39]. Photonic-circuit-integrated Ti:sapphire lasers and amplifiers demonstrate broad tunability from ~700 nm to 1000 nm, ultralow-threshold lasers and kilowatt level pulses reaching ~2 nJ energies, pumped directly by compact green diodes. These advances strongly motivate hybrid architectures in which an integrated Ti:sapphire gain chip is vertically or laterally coupled to nonlinear, CMOS-compatible Si$_3$N$_4$ waveguides[39,40] that convert the near-infrared output into an octave-spanning visible-to-infrared comb. Such an approach would combine the wide tunability and pulse quality of Ti:sapphire gain with the dispersion engineering and high Kerr nonlinearity of Si$_3$N$_4$, enabling fully integrated, diode-pumped, visible frequency comb engines[41].

The main obstacle on standard Si$_3$N$_4$ platforms is dispersion. In the spectral window between 800 nm and 850 nm, the material dispersion of Si$_3$N$_4$ and SiO$_2$ is strongly normal, and buried

Si$_3$N$_4$ waveguides remain in the all-normal regime even for sub-wavelength cross-sections. Figure 1 directly highlights this challenge. In Fig. 1(a) we illustrate the target nonlinear dynamics: a higher-order soliton, launched in the anomalous dispersion regime breathes and self-compresses, radiating dual dispersive waves in the normal dispersion regions on either side of the pump, thus seeding visible and near-infrared combs[17]. Figure 1(b) shows the calculated quasi-TE GVD for buried Si$_3$N$_4$ waveguides patterned on crack-free 400 nm LPCVD Si$_3$N$_4$ films and clad with SiO$_2$. In this geometry, the strong normal material dispersion of both core and cladding dominates, and the total GVD remains normal at the Ti:sapphire pump wavelength of 825 nm for all core widths considered. The dashed curves indicate the material dispersion of Si$_3$N$_4$ and SiO$_2$, highlighting that both are normal below ∼1.2 µm. In contrast, Fig. 1(c) shows the quasi-TE GVD for identical cores but left uncladded with air on the top and sides. The increased index contrast between Si$_3$N$_4$ (n ≈ 2.0) and air (n = 1) yields tighter mode confinement and substantially stronger geometric dispersion, which counteracts and eventually overcomes the normal dispersion contribution of the material. For core widths between 0.55 µm and 0.8 µm, the total dispersion is pushed into the anomalous regime at around 825 nm and exhibits two zero-dispersion wavelengths located in the visible and near-infrared.

This dispersion leverage is the analogue of what microstructured air–silica fibers achieved two decades ago, where the introduction of air increased confinement, rendered the total dispersion anomalous near 800 nm, and enabled the first Ti:sapphire pumped supercontinua[42] and later studied in suspended waveguides[43,44], except that here everything is defined by standard lithography on a chip. The finite-element simulations in Fig. 1(b,c) thus show that simple air-cladding of CMOS-foundry Si$_3$N$_4$ films suffices to open a tunable anomalous window from about 650 to 1050 nm across the full tuning range of Ti:sapphire lasers. In such a dual zero-dispersion wavelength landscape, the span of the supercontinuum is largely determined by the locations of the short- and long-wavelength dispersive waves in the normal-dispersion regions. Their central frequencies are set by the phase-matching between the soliton and a linear wave[45]:

$$\Delta\beta(\omega_{DW}) = \beta(\omega_{DW}) - \beta(\omega_s) - \frac{\omega_{DW} - \omega_s}{v_g} - \frac{\gamma P_s}{2},$$

where $\beta(\omega)$ is the propagation constant, $\omega_s$ and $\omega_{DW}$ are the soliton and dispersive-wave angular frequencies, $v_g$ is the soliton group velocity and $P_s$ its peak power. The wavelength of dispersive-wave is located at the zero-crossing of the phase-mismatch $\Delta\beta$. Figure 2(a) shows $\Delta\beta$ for our air-cladded waveguides as a function of wavelength and core width, assuming a soliton centered at 825 nm with 150 pJ pulse energy in the quasi-TE mode. Two distinct phase-matching points emerge, a visible dispersive wave tunable from 512 nm to 684 nm and a near-infrared dispersive wave tunable from 1164 nm to 1354 nm as the core width is increased from 0.55 µm to 0.8 µm. The reduction in spectral separation between the 825 nm soliton and the visible dispersive wave, compared to telecom wavelength-pumped platforms, is evident in these curves and directly reflects the reduced phase-mismatch barrier when pumping closer to the visible.

We experimentally verify this by seeding the air-cladded Si$_3$N$_4$ waveguides with an 80 MHz mode-locked Ti:sapphire laser centered at 825 nm, delivering 130 fs pulses. The waveguides are fabricated on commercially available, crack-free 400 nm LPCVD Si$_3$N$_4$ films on thermally oxidized silicon using subtractive electron-beam lithography. ma-N 2400 resist is spin-coated, patterned, and developed, and the Si$_3$N$_4$ is etched using CHF$_3$/O$_2$ reactive-ion etching. After resist stripping and cleaning, the chips are cleaved to form edge-coupling facets. The resulting waveguides exhibit near-vertical sidewalls and low roughness, as confirmed by scanning electron microscope (SEM) images in Fig. 2(d–f). We restrict the nonlinear sections to 7 mm of uniform, air-cladded waveguide to avoid additional dispersion and nonlinearity in the coupler regions. Propagation loss is measured to be ~3.6 dB/cm at 825 nm with the cut-back method. The coupling losses are about 8 dB per facet using 0.6 NA aspheric lenses, dominated by the mode mismatch between the free-space Gaussian beam and the tightly confined quasi-TE waveguide mode.

Figure 2(b) shows the measured output spectra for a set of air-cladded waveguides with core widths between 0.55 µm and 0.8 µm, pumped with 150 pJ coupled pulse energy in the quasi-TE mode. In all cases, we observe dual dispersive waves whose central wavelengths tune monotonically with core width, in excellent agreement with the phase-matching curves of Fig. 2(a). For the narrowest, 0.55-µm-wide waveguide, the 30 dB spectral span extends from 490 nm to 1088 nm, while for a 0.8-µm-wide waveguide the span extends from 583nm to 1331 nm. Both the short- and long-wavelength edges red-shift as the width increases, as indicated by dashed arrows and in agreement with phase-matching condition. Figure 2(c) compares the spectra for intermediate widths of 0.6 µm and 0.75 µm, which exhibit 30 dB spans from 493 nm to 1176 nm and 554 nm to 1316 nm, respectively, both well beyond a full octave. Thus, by tuning a single geometric degree of freedom, we realize continuously tunable visible dispersive waves while maintaining octave-spanning coverage into the near-infrared. The manifestation of this tunability is captured in Fig. 2(g). Here, we image the visible light scattered from the waveguides from the top while the invisible Ti:sapphire pump is launched from the left. The color of the scattered light shifts across the chip as the core width varies, directly visualizing the tunable visible dispersive wave. For each device, the indicated wavelength corresponds to the measured peak of the visible DW near its generation threshold. The data in Fig. 2 confirms that air-cladded $Si_3N_4$ waveguides support efficient, widely tunable visible comb generation when driven by modest-energy Ti:sapphire pulses at 825 nm, and that the DW positions and bandwidth are accurately captured by analyzing the phase matching.

A key advantage of the air-cladded geometry is its strong polarization-dependent dispersion, which allows the same physical waveguide to be operated either in an all-normal or anomalous-dispersion regime by rotating the input polarization. The guided modes experience a highly asymmetric cladding environment with $SiO_2$ (n ≈ 1.45) beneath and air (n = 1.0) above and laterally. As a result, quasi-TE and quasi-TM modes sample the core and cladding differently and exhibit a significantly different dispersion landscape. Figure 3(a) shows the measured near-saturated spectra for both polarization states at 150 pJ coupled energy in a 0.7-µm-wide air-cladded waveguide, together with the calculated GVD for each mode. The quasi-TE mode resides in the anomalous dispersion regime at 825 nm, between two zero-dispersion wavelengths, whereas the quasi-TM mode remains in the all-normal regime across all widths. Correspondingly, the quasi-TE spectrum spans 742 nm in 30 dB bandwidth, exhibiting a characteristic two-step structure associated with dual dispersive waves, while the quasi-TM spectrum broadens only to 159 nm and remains nearly symmetric around the pump.

Figures 3(b) and 3(c) map the evolution of the output spectra with increasing pulse energy for the quasi-TE and quasi-TM modes, respectively. At low energies (< 20 pJ), both polarizations show similar, modest broadening dominated by weak SPM, consistent with the limited nonlinear phase accumulation. As the energy increases, a strong asymmetry emerges. In the quasi-TM mode, the spectrum broadens slowly and symmetrically, indicative of all-normal dispersion where SPM and GVD act in the same direction. In contrast, the quasi-TE mode bandwidth grows rapidly once the soliton order exceeds unity, and two clear spectral steps develop as the visible and near-infrared dispersive waves are seeded at slightly different thresholds, as highlighted by the 30 dB bandwidth inset in Fig. 3(b). Importantly, the quasi-TE spectrum reaches a 159-nm-wide bandwidth already at ~35 pJ, whereas the quasi-TM mode requires the full 150 pJ to reach the same span. This polarization-dependence directly reflects the efficiency gained by anomalous dispersion and soliton-driven dynamics at short wavelength.

To assess the comb structure of the generated dispersive waves, we isolate the dispersive wave with a long-pass filter and detect it with a photodiode. The resulting radio-frequency beat note, shown in Fig. 3(d) for the quasi-TE mode at 150 pJ, exhibits a signal-to-noise ratio of 65 dB at the laser repetition rate, confirming that the dispersive wave inherits the comb structure of the pump[24]. Given the octave-spanning bandwidth, such visible-to-infrared supercontinua provide the ingredients for self-referenced frequency combs[15]. The onset of dispersive-wave generation and octave coverage can be estimated by the soliton fission length $L_{fiss} = L_D/N$, which approximates the position of maximum compression of a higher-order soliton[17,45], where $L_D = T_0^2/\beta_2$ is the dispersion length for a pulse with characteristic duration $T_0$ and $N = \sqrt{L_D/L_{NL}}$ is the soliton number with nonlinear length $L_{NL} = 1/(\gamma P_s)$. For our 7.1 mm-long waveguides, the calculated fission length as a function of pulse energy is overlaid in Fig. 3(e) on images of the visible light scattered along the device. We find that visible dispersive-wave generation sets in when $L_{fiss}$ becomes comparable to the physical length, corresponding to on-chip pulse energies in the range 30 pJ to 60 pJ depending on core width. This analysis suggests straightforward pathways to further reduce the energy threshold, by increasing the waveguide length, using shorter input pulses, or employing dispersion alternating waveguides that enhance soliton self-compression[46].

Finally, we demonstrate that the same dispersion-engineered, air-cladded $Si_3N_4$ waveguides efficiently extend to the GHz repetition-rate regime, which is highly relevant for astrocombs and broadband metrology. Figure 4(a) sketches our experimental setup. A compact, three-element Ti:sapphire Kerr-lens mode-locked oscillator, pumped directly by a single green diode, delivers 50 fs pulses at 810 nm with a repetition rate of 1 GHz and pulse energies up to 150 pJ. The cavity design follows recent work on diode-pumped Ti:sapphire GHz oscillators[47], achieving average powers of 180 mW. We couple these GHz pulses via aspheric lenses into the quasi-TE mode of a 0.6-µm-wide air-cladded $Si_3N_4$ waveguide. Despite the 12.5× higher repetition rate and correspondingly lower available pulse energies compared to typical MHz-class systems, on-chip pulse energies of only 38 pJ suffice to generate an octave-spanning spectrum, as shown in Fig. 4(b). The input spectrum is plotted in gray, while the output shows octave coverage from visible to infrared. This result highlights the efficiency of our dispersion-engineered waveguides, the combination of short-wavelength pumping, tight mode confinement, and anomalous GVD allows us to reach octave bandwidths at GHz repetition rates with pulse energies that are readily delivered by diode-pumped Ti:sapphire oscillators.

In conclusion these results establish air-cladded, dispersion-engineered $Si_3N_4$ waveguides as an effective bridge between emerging integrated Ti:sapphire gain platforms and on-chip octave visible frequency combs. By pumping in the 800–830 nm range, we leverage the broadband gain and ultrashort pulses of Ti:sapphire while reducing the phase-mismatch barrier to visible dispersive waves and exploiting the enhanced nonlinearity at short wavelength. Air-cladding provides a simple but powerful means to overcome the strongly normal material dispersion of $Si_3N_4$ and $SiO_2$, opening a tunable anomalous window from 650 nm to 1050 nm across the Ti:sapphire tuning range, and enabling dual dispersive-wave generation with full octave coverage from 490 nm to 1330 nm at on-chip pulse energies in the tens of picojoules. Efficient GHz octave coverage is demonstrated directly driven by a compact single-diode-pumped Ti:sapphire oscillator. The strong polarization dependence of the dispersion landscape further offers a unique degree of freedom by simply rotating the input polarization, the same waveguide can be switched between an all-normal regime suitable for highly coherent, flat, compressible spectra and an anomalous, soliton-dominated regime that efficiently seeds visible dispersive waves and octave combs. Looking ahead, hybrid integration of Ti:sapphire gain media with $Si_3N_4$ nonlinear waveguides is already being explored in heterogeneous Ti:SaOI–$Si_3N_4$ amplifier schemes. Our results directly align with these efforts where the 400 nm $Si_3N_4$ films used for our air-cladded waveguides can be deposited onto Ti:SaOI substrates, and compact vertical couplers can route the integrated Ti:sapphire oscillator or amplifier output into dispersion-engineered nonlinear sections on the same chip[37–39]. Such a platform would enable fully integrated, diode-pumped, octave-spanning visible frequency combs with GHz repetition rates, tailored for environmental monitoring, astronomical spectrograph calibration, and portable precision metrology. Beyond supercontinuum generation, the anomalous dispersion landscape and strong confinement at short wavelength should also be beneficial for visible soliton microcombs, $\chi^3$-based parametric oscillators, and nonlinear microscopy on a chip. The combination of integrated Ti:sapphire sources and air-cladded $Si_3N_4$ nonlinear waveguides thus points toward a new class of compact, high-performance visible photonic engines that unite the best of bulk ultrafast lasers with the scalability and robustness of integrated photonics.

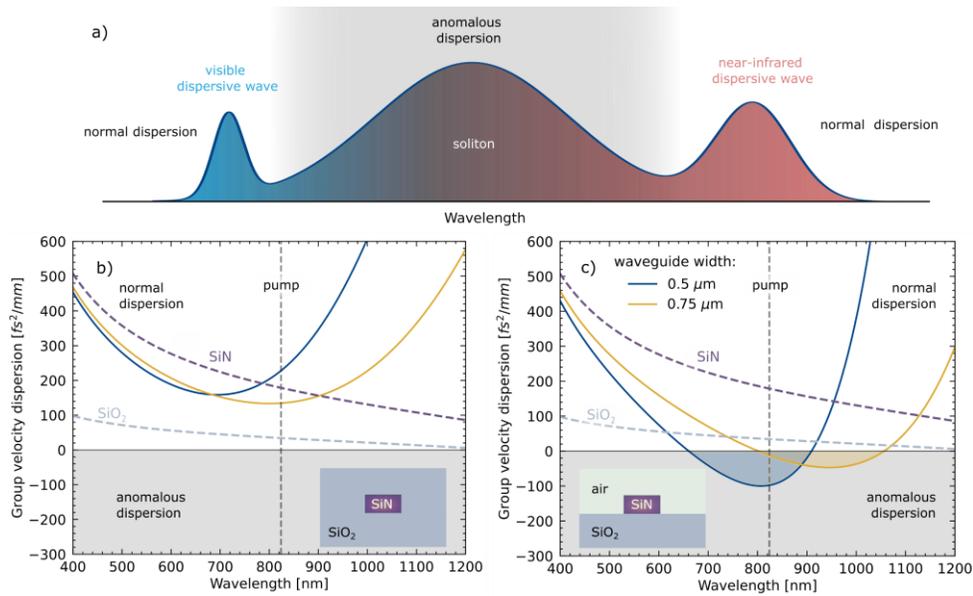

Fig. 1. **Dispersion-engineered air-cladded silicon-nitride waveguides for broadband visible-to-near-infrared frequency-comb generation.** a) Conceptual schematic of broadband comb generation: an excited higher order soliton in the anomalous dispersion regime breathes, and seeds dual visible and near-infrared dispersive waves phase-matched in the normal dispersion regime. b) Quasi-TE mode group velocity dispersion of buried waveguides patterned on crack-free 400-nm-thick silicon-nitride films, where strong material dispersion dominates resulting in all normal dispersion at the pump wavelength 825 nm (vertical dashed line). In contrast, air-cladded waveguides in c), where the higher refractive index contrast between core and upper-cladding results in tight confinement and strong geometrical dispersion, pushing the total dispersion to the anomalous dispersion regime with two zero-dispersion wavelengths. The material dispersion for both silicon-nitride and silicon-dioxide are shown in dashed lines for comparison.

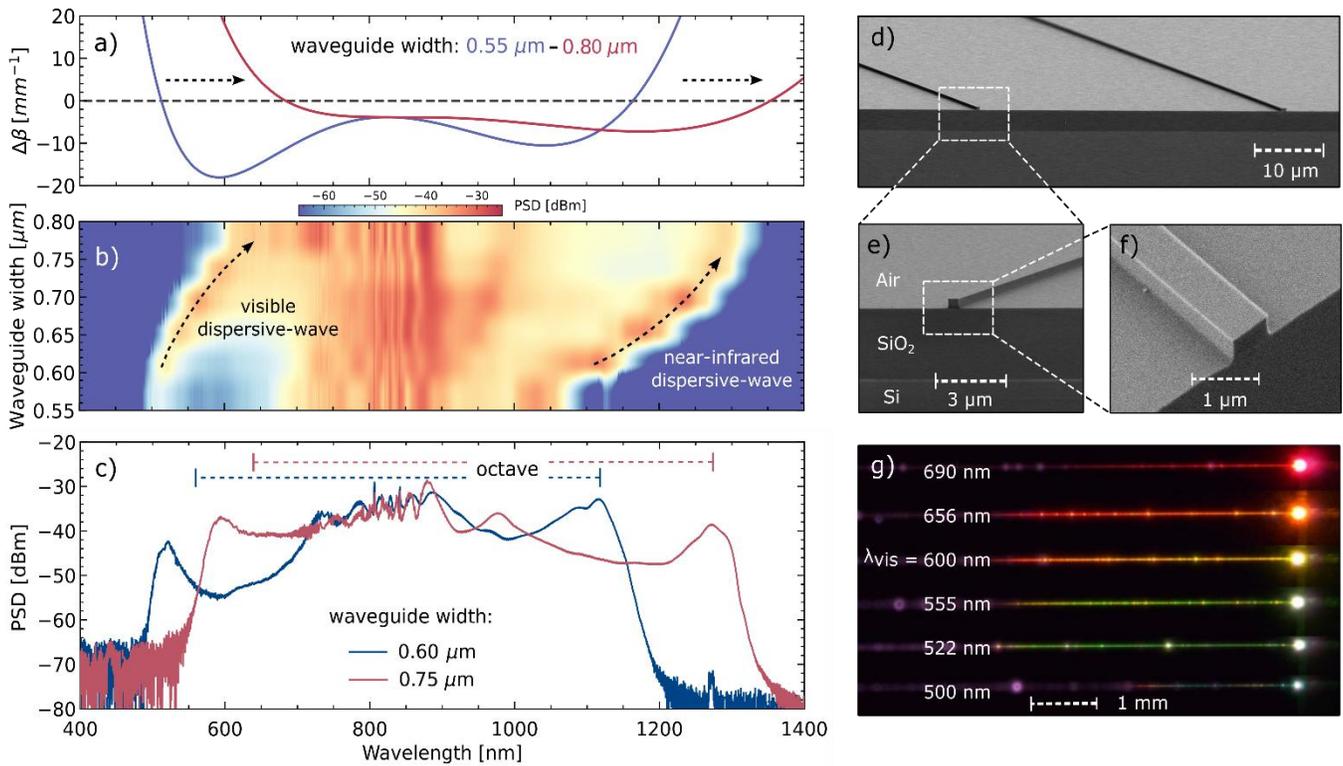

Fig. 2. **Tunable broadband visible frequency comb generation.** a) Calculated phase-mismatch parameter Δβ between a soliton at pump wavelength 825 nm and linear wave for air-cladded waveguides with quasi-TE mode. Two dispersive waves appear at the locations of zero phase-mismatch. b) Experimentally observed broadband visible comb when increasing the waveguide width from 0.55 µm to 0.8 µm (pulse energy 150 pJ). Both visible and near-infrared dispersive waves are observed experiencing red-shift with increasing waveguide width, as indicated by dashed arrows. All measured spectra have octave coverage. c) Output power spectral density (PSD) for two waveguide widths for waveguide width of 0.6 µm with 30dB span from 493 nm to 1176 nm in comparison to waveguide width of 0.75 µm with span from 554 nm to 1316 nm . d-f) Scanning electron microscope images of the air-cladded waveguides with different magnifications. g) The tunable visible light scattered from the waveguides is captured from the top by a camera where the invisible pump is launched from left. $\lambda_{vis}$ indicates the measured peak of the dispersive wave close to threshold.

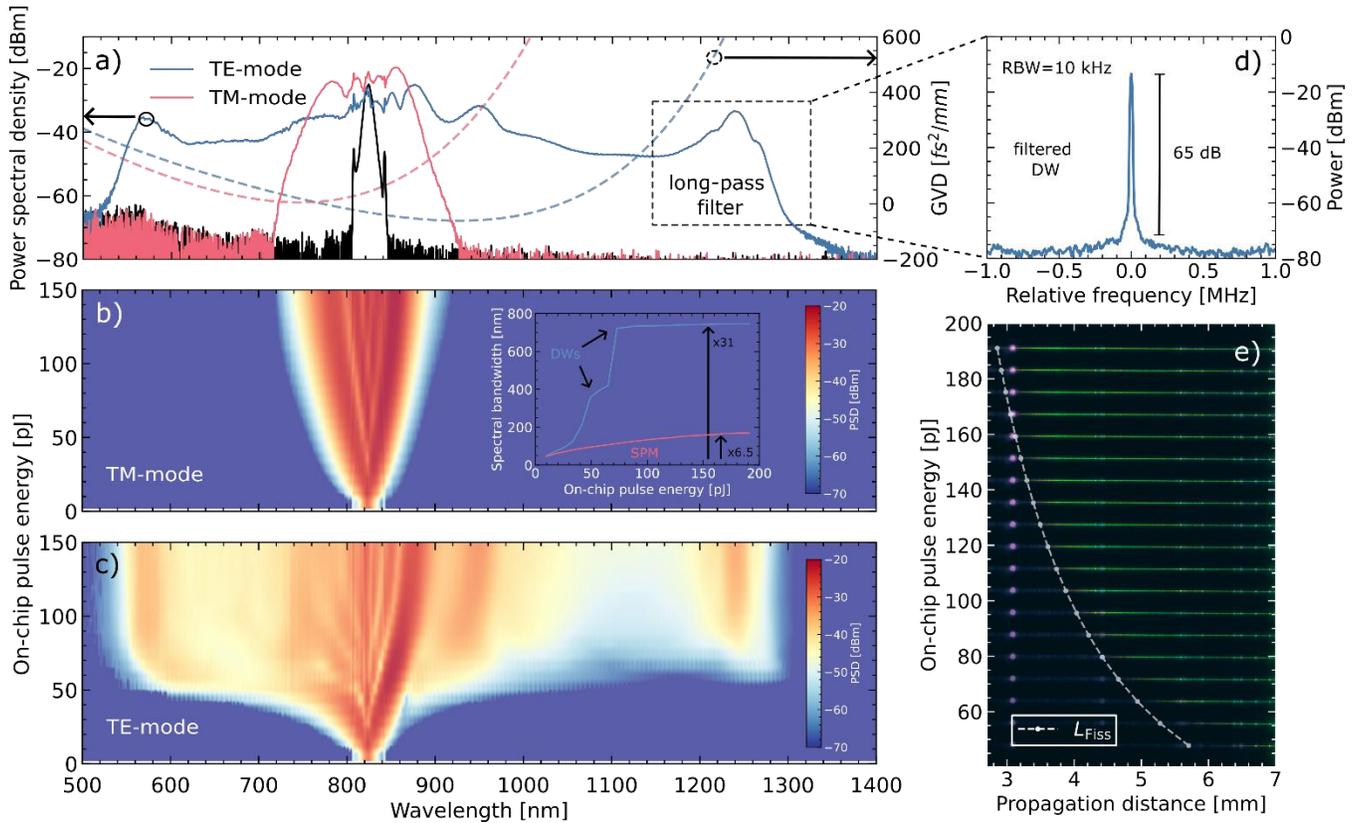

Fig. 3. **Asymmetrical polarization-dependence of air-cladded waveguides on dispersion regimes from all-normal to anomalous dispersion.** (a) Measured saturated output spectra for quasi-TE and TM modes of air-cladded waveguide with 0.7 µm core-width at 150 pJ pulse energy. The right axis shows calculated group velocity dispersion for both modes, highlighting anomalous (normal) dispersion at pump wavelength for quasi-TE mode (quasi-TM mode). The pump spectrum is shown in black. (b) & (c) Measured pulse energy influence on output spectra for both modes. The quasi-TM mode shows slow increase of spectral bandwidth from 24 nm to 159 nm dominated by self-phase modulation (SPM) in the all-normal dispersion regime in contrast to rapidly increasing bandwidth from 24 nm to 742 nm in the anomalous dispersion regime for quasi-TE mode, dominated by efficient soliton dynamics and dual dispersive wave generation (DWs). The inset shows the measured 30 dB bandwidth for both modes. (d) Beatnote of the dispersive wave for quasi-TE mode at 150 pJ pulse energy. (e) Captured visible dispersive wave generation for different pulse energies. The dashed line represents the calculated soliton fission length ($L_{Fiss}$).

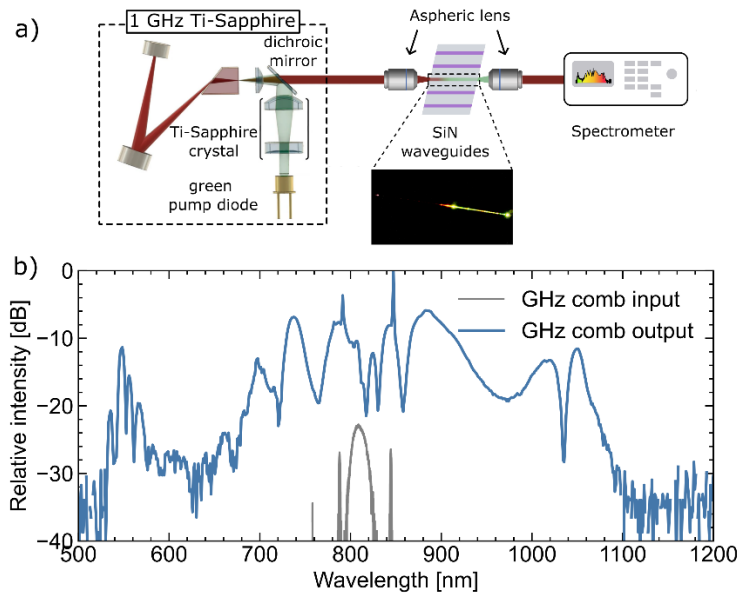

Fig. 4. **Coherent octave 1 GHz frequency comb generation.** A compact single-diode-pumped Ti-Sapphire oscillator directly drives the octave generation from air-cladded silicon-nitride waveguides. (a) Schematic of the experimental setup for GHz frequency comb generation. A three-element diode-pumped Kerr-lens mode locked Ti-Sapphire oscillator delivers 50 fs pulses with up to 150 pJ energy at 810 nm. An aspheric lens is used to launch the GHz pulses into the quasi-TE mode of air-cladded silicon-nitride waveguide with 0.6 µm core-width. The inset shows a photograph of the visible light generated. (b) Measured input and output spectra.

**Funding.** National Institutes of Health (123456); National Science Foundation (456789).

**Acknowledgement.** Munich Quantum Valley Project TeQSiC, DFG Project 541267874, The German Federal Ministry of Research, Technology and Space, Quantum Systems, 13N17314, 13N1734. Engineering and Physical Sciences Research Council (EP/Y011422/1); Science and Technology Facilities Council (ST/V000403/1, ST/Y001273/1, ST/X004503/1); Innovate UK (10073639). A.A. acknowledges funding from Max Planck School of Photonics and support from TDSU department at MPL Erlangen.

**Disclosures.** The authors declare no conflicts of interest.

**Author Contributions.** A.A. conceived the study and wrote the manuscript, with input from all authors. A.A. designed, modelled, fabricated the waveguides and performed the experiment with a commercial laser source. H.O., E.A., P.C.M. and D.T.R. constructed the GHz source and measurement. D.T.R. and P.D.H. supervised the project.

**Data Availability.** Data underlying the results presented in this paper are not publicly available at this time but may be obtained from the authors upon reasonable request.